# Observing Outer Planet Satellites (except Titan) with JWST: Science Justification and Observational Requirements

Laszlo Keszthelyi[1], Will Grundy[2], John Stansberry[3], Anand Sivaramakrishnan[3], Deepashri Thatte[3], Murthy Gudipati[4], Constantine Tsang[5], Alexandra Greenbaum[6], Chima McGruder[7]

[1]U.S. Geological Survey, Astrogeology Science Center, 2255 N. Gemini Dr., Flagstaff, AZ 86001. (laz@usgs.gov)

[2]Lowell Observatory, 1400 W. Mars Hill Rd., Flagstaff, AZ 86001. (grundy@lowell.edu)

[3]Space Telescope Science Institute, 3700 San Martin Dr., Baltimore, MD 21218. (jstans@stsci.edu; thatte@stsci.edu; anand@stsci.edu)

[4]California Institute of Technology, Jet Propulsion Laboratory, 4800 Oak Grove Dr., Pasadena, CA 91109. (gudipati@jpl.nasa.gov)

[5]Southwest Research Institute, Department of Space Studies, 1050 Walnut St., Suite 300, Boulder, CO 80302. (con@boulder.swri.edu)

[6] Department of Physics and Astronomy, 3400 N. Charles Street, Baltimore, MD 21218. (agreenba@pha.jhu.edu)

[7]Department of Physics & Astronomy, 1408 Circle Drive, Knoxville, TN 37996. (cmcgrud1@vols.utk.edu)

ABSTRACT

The James Webb Space Telescope (JWST) will allow observations with a unique combination of spectral, spatial, and temporal resolution for the study of outer planet satellites within our Solar System. We highlight the infrared spectroscopy of icy moons and temporal changes on geologically active satellites as two particularly valuable avenues of scientific inquiry. While some care must be taken to avoid saturation issues, JWST has observation modes that should provide excellent infrared data for such studies.

KEYWORDS

Solar System
Astronomical Techniques

## 1. Introduction

Starting in 2018, the James Webb Space Telescope (JWST) is expected to provide a groundbreaking new tool for astronomical observations, including planetary satellites as close as Phobos and Deimos. The purpose of this paper is to identify the types of observations of outer planet satellites that might be obtained using JWST and to provide suggestions to help successfully complete such observations.

We note that Titan is covered in another paper in this special issue [Nixon et al., 2015] and is thus not discussed here. Similarly, we do not repeat material covered by other papers focused on satellites associated with planetary rings or other small bodies that can have satellites [Thomas et al., 2015; Rivkin et al., 2015; Tiscareno et al., 2015; Kelley et al, 2015; Santos-Sanz et al., 2015; and Parker et al., 2015].

## 2. Relevant JWST Capabilities and Constraints

Colloquially, JWST is referred to as the next Hubble Space Telescope (HST). However, JWST is an infrared space observatory, with scientific strengths notably different from HST. Spatial resolution can be slightly better than HST but the design of JWST is really optimized for obtaining infrared observations of faint astronomical targets. It will fly with 4 instruments: the near-IR camera (NIRCam), the near-IR spectrograph (NIRSpec), the near-IR imager and slitless spectrograph (NIRISS), all operating in the 0.6 – 5.0 µm range, and the 5.0-29 µm mid-IR imaging spectrometer (MIRI) [Gardner et al., 2006; Milam et al., 2015]. While all four instruments can in principal acquire data simultaneously, for observations of satellites they will be operated individually so that the target can be placed into the selected instrument field of view. For observations of the outer-planet satellites the overall combination of spatial and spectral resolution rivals that of most previous spacecraft data (Table 1).

[TABLE 1]

The strengths of JWST do come at a price. For example, JWST will not observe in the ultraviolet region where Hubble Space Telescope (HST) observations of Solar System targets generated major discoveries. Also, the emphasis on faint targets means that there are some observation modes where instruments will saturate when looking at larger outer planet satellites. One of the most significant constraints on JWST operations is the need to maintain the orientation of the sunshield. This limits observations within the Solar System to around the time when the target is near quadrature (solar elongation limits of 85° – 135°). In other words, observation opportunities come approximately twice a year. Table 2 lists the dates when different outer planets (and hence their satellites) are observable by JWST during its nominal mission.

[TABLE 2]

## 3. Suggested Observations

Based on the strengths and constraints of JWST, we suggest two primary science themes: (1) completing the infrared survey of major satellites and (2) monitoring surface changes of geologically active satellites.

*3.1 Spectroscopy of outer planet satellites*

The JWST instrument suite provides a unique opportunity to obtain high spectral resolution infrared spectra from planetary satellites in wavelength regions that cannot be observed from Earth. For most Galilean and Saturnian satellites, *Galileo* NIMS and *Cassini* VIMS obtained spectra out to ~5 μm at moderately high spectral resolution and at much higher spatial resolutions than JWST can achieve (Table 1). At longer wavelengths, the *Cassini* CIRS instrument has obtained data at spectral resolutions comparable to JWST. Voyager IRIS instruments were able to collect longer IR data at spectral resolution not much less than the JWST instruments but their spatial resolution was very low. Thus the *type* of data JWST could collect for Solar System satellites is not unique or unprecedented.

However, the data collected by flyby and orbiter missions is incomplete and/or collected under illumination conditions that are less than desirable for spectroscopy (e.g., high phase angles). Also, most of the satellites of Uranus and Neptune were either undiscovered or undetectable by Voyager IRIS. The lack of a complete and consistent set of infrared spectra is a significant obstacle for systematic or comparative studies of satellites across our Solar System. Such studies hold out a strong potential for major advances in our understanding of the formation and evolution of the Solar System and key chemical building blocks for biologic processes [e.g., Dalton et al., 2010].

JWST's key scientific contribution could be determining the compositions of irregular satellites. Even at very low spatial resolution, near-IR spectroscopy is sensitive to $H_2O$ and other ices, as well as silicates and spectral slopes characteristic of complex organic "tholins" [e.g., Dalton et al., 2010]. A strong absorption band around 3.1 μm is an especially sensitive tracer of $H_2O$, both in the form of ice and also structurally bound up in minerals. Other, more volatile ices such as $CO_2$, $NH_3$, and $CH_4$ have comparably strong and readily distinguishable bands in the 2.5 to 5 μm range (Fig. 1). This is also a region with important spectral features for organic compounds [e.g., NRC, 2007]. Atmospheric absorption makes these wavelengths extremely challenging to observe from Earth, especially for such faint targets, but JWST has the sensitivity to provide unique compositional data on irregular satellites. For example, in the 1-2.5 micron region of the NIR, amorphous vs. crystalline surface composition of icy bodies could be surveyed extensively using JWST NIRSpec.

[FIG 1]

We note that additional laboratory spectra are needed to fully utilize the new JWST data. This is because the spectral regions that cannot be observed from Earth (but which can be observed by JWST) have received relatively little attention from the groups that build spectral libraries and other essential reference data. The appropriate data can be obtained from existing laboratories with modest modification of detectors and no significant change to data processing and distribution methods [e.g., de Berg et al., 2008].

We expand upon one of many research problems that could be addressed with spectroscopy obtained by JWST of outer planet satellites. In addition to their large, regular satellites, the four giant planets host retinues of irregular satellites. These are small bodies captured into wide, inclined, and eccentric orbits. They offer opportunities to sample the various planetesimal populations they were captured from, and with over a hundred known, their numbers enable statistical studies that could not be done for an individual object. Cassini explored the Saturnian irregular satellite Phoebe, finding abundant ices including volatile $CO_2$, coupled with a rock-rich density like that of Pluto. These characteristics were interpreted as suggesting kinship with the relict planetesimals in the Kuiper belt [Johnson and Lunine, 2005]. But comparably-sized Kuiper belt objects (KBOs) have much lower bulk densities [e.g., Stansberry et al., 2012, Vilenius et al., 2014], as do comets thought to be collisional fragments from the Kuiper belt. This raises the question whether Phoebe is a collisional fragment from (a) a denser, Pluto-sized KBO or (b) from a planetesimal sub-population not well represented among KBOs or (c) has subsequent evolution increased its density (e.g., Castillo-Rogez et al., 2012)? Studies of the ensemble population could address many open questions about how and when the irregular satellites were captured, what planetesimal populations they represent, and how they subsequently evolved.

Comparisons between the compositions of irregular satellites and KBOs would be valuable. This could be in the context of an experiment designed to try to link both irregular satellites and KBOs to their ancestral planetesimal populations. Alternatively, it could proceed from the assumption of shared initial compositions to investigate potential space-weathering effects by comparing KBOs at 40+ AU from the Sun with irregular satellite populations at 5, 10, 20, and 30 AU. Irregular satellites occur in clusters in orbital element space (Fig. 2), thought to originate from breakups of parent bodies, and possibly related to their capture (e.g., Nicholson et al., 2008). It would be useful to compare the compositions of members of irregular satellite clusters to see if they are homogeneous. Differences could provide clues to size-dependent space weathering effects, to differentiation of the progenitor into distinct core and mantle compositions, or to an as-yet unidentified dynamical mechanism that concentrates dissimilar satellites into specific regions of orbital phase space.

Irregular satellites are also important sources of dust in the giant planet systems. Dust orbits evolve under effects of radiation pressure and solar tides [e.g., Tamayo et al., 2011]. The dust can then form rings [e.g., Verbiscer et al., 2009], be swept up by larger satellites [e.g., Buratti et al., 2005], and feed neutral gas tori via sputtering with gas molecules eventually becoming ionized loading the magnetospheres or being removed by the solar wind. By linking the the sizes, densities, and albedos of dust particles to the source satellite surface compositions, JWST can bring new insight into the role of satellites in the complex dance of dust particles in these miniature solar systems.

[FIG 2]

We suggest that there is special value in JWST obtaining global spectra (~4 hemispheric views) from all significant planetary satellites under consistent high-quality (i.e., high-sun) illumination. JWST observations would greatly benefit from laboratory studies needed to interpret such data.

*3.2 Detecting geologic activity on planetary satellites*
As our spacecraft have spread across our Solar System, we have discovered that many of the outer planet satellites are remarkably active. Io, Triton, and Enceladus have active eruptions [Morabito et al., 1979; Soderblom et al., 1990; Hansen et al., 2006]. The recent suggestion of active plumes above Europa is especially exciting because it may provide samples from a habitable environment that is otherwise extremely challenging to access [Roth et al., 2014]. However, eruption plumes are best detected in the ultraviolet, a region of the electromagnetic spectrum that JWST cannot observe.

Instead, JWST will be able to detect changes on the surface that are indicative of temporal variations in composition and/or temperature. However, this requires baseline observations to compare against. A benefit of the restricted range of illumination and viewing geometries of JWST for observing outer planet satellites is that radiometric and photometric corrections will be minor. For measuring thermal emission, it is somewhat preferable to obtain data without the influence of direct sunlight. However, due to the viewing geometry JWST can observe outer planet satellites in eclipse only when the target is very close to the bright primary that it orbits. Similarly, only a thin sliver of the night side will be visible with challenging stray light issues from the adjacent sunlit surface. While longer wavelengths are not significantly affected by reflected sunlight, the inability to observe the diurnal temperature variations makes it difficult to study some thermo-physical parameters such as thermal inertia.

The target where it is certain that JWST can observe significant changes is Io. The observations every ~6-months that JWST can make of the Jovian system is very well suited for monitoring the creation and fading of colorful plume deposits on Io which typically happen on a timescales of several months and have diameters of many hundreds of kilometers (Fig. 3) (McEwen et al., 1998). Simulation of JWST observations illustrate how effectively the major volcanoes can be spatially resolved (Fig. 4). The JWST wavelengths are also particularly well-suited for providing invaluable new constraints on the largest unresolved scientific problem in the post-Galileo era: pinning down the eruption temperature of Io's lavas. While it is certain that silicate volcanism is extremely prevalent on Io, the uncertainty in the eruption temperature is $>\pm 100$ K and hence the composition of the lavas is poorly constrained (Keszthelyi et al., 2007). This in turn leaves great uncertainty about the composition and state of Io's mantle, which cascades into uncertainty about how tidal heating works in the Jovian system.

[FIG 3]

[FIG 4]

JWST's ability to measure Io's infrared emission at an unprecedented spectral range and resolution could play a key role in solving this problem. The emission from a volcanic eruption is the sum of the contributions from surfaces at a continuum of temperatures ranging from the eruption temperature to ambient (e.g., Davies, 1996). The hottest surfaces cool extremely quickly (of order 100 K/s) and therefore make up only a very small fraction of the area on a volcano. This is only partially offset by the fact that thermal emission depends on temperature to the fourth power. In order to isolate this small hottest component of thermal emission, it is essential to obtain relatively high spectral resolution in the short to near IR. This is because it is the steep short wavelength side of the Planck Function that is most sensitive to the high temperature component (Fig. 5). Furthermore, the models used to estimate the continuum of lava surface temperatures are best constrained if the data covers a wide region of the infrared, including data beyond 10 microns, at moderate spectral resolution (e.g., Davies et al., 2010). Furthermore, if the observations have sufficient spectral resolution in the 5-10 micron region, these same models can be used to detect and quantify the role of lower temperature sulfurous volcanism that is present on Io but has been essentially unobservable by most instruments (Lopes et al., 2001). JWST's instruments are particularly well suited to obtaining these kinds of data. In addition, mapping the passive surface temperatures at these long wavelengths on Io will greatly enhance our understanding of the vapor pressure equilibrium of $SO_2$ surface frost and the extremely tenuous atmosphere. This relationship is poorly understood but is known to be a strong function of the surface temperature. The interplay between Io's $SO_2$-dominated frost and atmosphere is likely to provide new insight into the processes important on other "airless" icy bodies.

[FIG 5]

There are some subtleties in the data acquisition that will have to be considered. As discussed in the following section, selecting the wrong observation modes can lead to issues with saturation, especially at the shorter wavelengths where reflected sunlight is important. For volcanism on Io, the timescale at which emission varies is closely linked to the wavelengths, with changes being more rapid at shorter wavelengths. Thus taking data more rapidly in the NIR and SWIR than in the thermal IR and beyond is beneficial.

We suggest that there is considerable scientific merit in JWST obtaining multiple observations of each of the geologically active planetary satellites. JWST has the spatial resolution, wavelength coverage, and spectral resolution to detect significant geologic events on the outer planet satellites. Ideally, such data would be collected at each opportunity (biannually) to detect large-scale changes, if they occur, and maintain a current baseline if no changes occurred.

**4. Action Items**

These two types of JWST observations will enable compelling science of outer Solar System satellites. However, there are two technical concerns related to JWST itself and two other concerns about the infrastructure to support the analysis of the JWST data. Each of these issues is detailed in other publications, so only a brief summary is provided here.

*4.1 Saturation*

JWST is designed to observe extremely faint astronomical targets, so saturation is expected at some wavelengths when observing the planets in our Solar System [Norwood et al., 2015]. At Mars, the extremely low albedo and relatively small size of Phobos and Deimos should mitigate this problem [Villanueva et al., 2015]. Conversely, the high albedo and well-resolved nature of the Galilean satellites guarantee saturation at shorter wavelengths in the most sensitive observation modes. Similar concerns extend to the larger Saturnian satellites. During major volcanic eruptions the thermal emission from Io can readily exceed the brightness from reflected sunlight, so saturation is a serious concern across much of the spectrum. Saturation concerns only drop away at Uranus and Neptune. However, the initial analysis indicates that there are observing modes for at least NIRCam that effectively avoid most saturation issues when observing even the Jovian and Saturnian satellites (Fig. 6). The other instruments should have lesser concerns related to saturation from reflected sunlight. While further analysis of saturation is clearly needed, we expect this to be solvable challenge for JWST observations of outer planet satellites.

[FIG. 6]

*4.2 Stray light*

Observing small faint objects close to large bright ones is always challenging. At Jupiter and Saturn, the larger satellites will regularly venture more than 100 pixels from the primary, which may be sufficient to limit the effect of stray light. At Uranus and Neptune, some of the satellites remain closer to the primary, so a detailed analysis of stray light is needed to provide quantitative requirements for JWST. The most serious stray light problems in optical systems are usually due to unpredicted issues, so adequate analysis of stray light may only be possible during later phases of integration and testing.

*4.3 Laboratory Spectra*

The translation of JWST observations into minerals and compounds will rely on comparisons with laboratory spectra of standard samples that have multiple components (ice, organics, and minerals). Lab data for Io's surface also need sulfur-containing molecules and ices at a wide variety of temperatures. Unsurprisingly, such spectra are relatively few for regions of the spectrum where the Earth's atmosphere is opaque. This is especially true at the cryogenic temperatures of the outer planet satellites. Given the time required to collect, process, and distribute these types of data, this should be a priority area of funding over the next few years.

*4.4 Cartography*

The ability to discern changes on the surfaces of the outer planet satellites requires the ability to precisely align data sets taken at different times. This requires precision cartography with the JWST observations projected into a consistent reference frame tied to the surface of the target bodies. This is not a capability often planned for astronomical observatories but the funding of appropriate software should not be neglected over the next few years.

**5. Conclusions**

JWST promises to be an important tool for studying planetary satellites. We encourage the observing community to use this document to formulate more specific observation plans. We encourage the JWST project to address the items in Section 4 in order to ensure the highest scientific return from outer planet satellite observations.

*Acknowlegments*
A. Sivaramakrishnan is supported via NASA Grant NNX11AF74G, as is A. Greenbaum, who also receives support through NSF Graduate Reseach Fellowship DGE- 123825. C. McGruder received support through the National Astronomy Consortium.

REFERENCES

Buratti et al., 2005. Spectrophotometry of the small satellites of Saturn and their relationship to Iapetus, Phoebe, and Hyperion. Icarus 175, 490-495.

Castillo-Rogez et al., 2012. Geophysical evolution of Saturn's satellite Phoebe, a large planetesimal in the outer Solar System. Icarus 219, 86-109.

Dalton, J. B., D. P. Cruikshank, K. Stephan, T. B. McCord, A. Coustenis, R.W. Carlson, and A. Coradini (2010) Chemical composition of icy satellite surfaces, Space Sci. Rev., v. 153, pp. 113-154.

Davies, A. G. (1996) Io's volcanism: Thermo-physical models of silicate lava compared with observations of thermal emission, Icarus, 124, 45-61.

Davies, A. G., (2010) The thermal signature of volcanic eruptions on Io and Earth, J. Volcanol. Geotherm. Res., 194, 75-99.

de Bergh C, Schmitt B, Moroz L.V. (2008) Laboratory data on ices, refractory carbonaceous materials, and minerals relevant to Transneptunian objects and Centaurs. In: Barucci MA et al (eds) The Solar System Beyond Neptune. The University of Arizona Press, Tucson, p. 483

Grundy, W. M., C. B. Olkin, L. A. Young, M. W. Buie, and E. F. Young, 2013, Near-infrared spectral monitoring of Pluto's ices: Spatial distribution and secular evolution, Icarus, v. 223, pp. 710-721.

Hansen, C. J., L. Esposito, A.I.F Stewart, J. Colwell, A. Hendrix, W. Pryor, D. Shemansky, R. West (2006) Enceladus' water vapor plume, Science, v. 311, pp. 1422-1425.

Johnson & Lunine, 2005. Saturn's moon Phoebe as a captured body from the outer solar system. Nature 435, 69-71.

Kelley, M.S. et al. (2015) PASP, this issue.

Keszthelyi, L., W. Jaeger, M. Milazzo, J. Radebaugh, A.G. Davies, K.L. Mitchell (2007) Estimates for Io eruption temperatures: Implications for the interior, Icarus, 192, 491-502.

Lopes, R. M. C., L. W. Kamp, S. Doute, W. D. Smythe, R. W. Carlson, A. S. McEwen, P. E., Geissler, S. W. Kieffer, F. E. Leader, A. G. Davies, E. Barbinis, R. Mehlman, M. Segura, J. Shirley, and L. A. Soderblom (2001) Io in the near infrared: Near-Infrared Mapping Spectrometer (NIMS) results from the Galileo flybys in 1999 and 2000, J. Geophys. Res., 106, 33053-33078.

Milam, S. N., J. A. Stansberry, G. Sonneborn, and C. Thomas (2015) The James Webb Space Telescope's plan for operations and instrument capabilities for observations in the Solar System, Publications of the Astronomical Society of the Pacific, this issue.

Morabito, L. A., S. P. Synnott, P. N. Kupferman, S. A. Collins, (1979) Discovery of currently active extraterrestrial volcanism, Science, 204, 972.

McEwen, A. S., and 12 co-authors (1998) Volcanism on Io as seen by Galileo SSI. Icarus, 135, 181-219.

National Research Council Task Group on Organic Materials in the Solar System, (2007) Exploring Organic Environments in the Solar System, National Academies Press, 124 pp.

Nicholson et al. 2008. Irregular satellites of the giant planets. In The Solar System Beyond Neptune, Univ. of Arizona Press, pp. 411-424.
Nixon, C. et al. (2015) PASP, this issue.
Norwood, J. et al. (2015b) PASP, this issue.
Parker, A. et al. (2015) PASP, this issue.
Rivkin, A. et al. (2015) PASP, this issue.
Roth, J. S., K. D. Retherford, D. F. Strobel, P. D. Feldman, M. A. McGrath, F. Nimmo (2014) Transient water vapor at Europa's southern pole, Science, 343, 171-174.
Santos Sanz, P. et al. (2015) PASP, this issue.
Soderblom L. A., S. W. Kieffer, T. L. Becker, R. H. Brown, A. F. Cook II, C. J. Hansen, T. V. Johnson, R. L. Kirk, E. M. Shoemaker (1990) Triton's geyser-like plumes: Discovery and basic characterization, Science, v., 250, 410-415.
Stansberry et al., 2012. Physical properties of transneptunian binaries (120347) Salacia-Actaea and (42355) Typhon-Echidna. Icarus 219, 676-688.
Tamayo et al., 2011. Finding the trigger to Iapetus' odd global albedo pattern: Dynamics of dust from Saturn's irregular satellites. Icarus 215, 260-278.
Thomas, C. et al. (2015) PASP, this issue.
Tiscareno, M. et al. (2015) PASP, this issue.
Verbiscer et al., 2009. Saturn's largest ring. Nature 461, 1098-1100.
Vilenius et al., 2014. "TNOs are Cool": A survey of the trans-Neptunian region X. Analysis of classical Kuiper belt objects from Herschel and Spitzer observations. Astron. & Astrophys. 564, A35.1-18.
Villanueva, G. et al. (2015) PASP, this issue.

TABLES

**Table 1. Comparison of JWST instruments to some past and future planetary spectrometers.** Past instruments are in orange, JWST in green, and future instruments in blue. Spectral resolution is given as resolving power (R) defined as $\Delta\lambda/\lambda$ where $\Delta\lambda$ is the smallest difference in wavelength that the instrument can resolve at a wavelength $\lambda$. Note that the best spatial and spectral resolution are often in different imaging modes and cannot be obtained concurrently.

| Instrument | Wavelength (µm) | Best Spectral Resolution (R) | Best IFOV (mas) | Largest FOV (arcsec) |
|---|---|---|---|---|
| HST ACS/WFC | 0.1-1.1 | ~250 | 34 | ≤202x202 |
| HST NICMOS | 0.8-2.5 | ~200 | 43 | ≤51.2x51.2 |
| Galileo NIMS | 0.7-5.2 | ~450 | N/A | N/A |
| Cassini VIMS | 0.3-5.1 | ~350 | N/A | N/A |
| Cassini CIRS | 7.2-100 | ~2000 | N/A | N/A |
| Voyager IRIS | 2.5-55 | ~500 | N/A | N/A |
| NIRCam (Imaging - Short) | 0.6-2.3 | ~100 | 32 | 132X164 |
| NIRCam (Imaging - Long) | 2.4-5.0 | ~100 | 65 | 132X164 |
| NIRSpec (prism) | 0.6-5.0 | 100 | 100 | 1.6x1.6 |
| NIRSpec (grating) | 1.0-5.0 | 2700 | 100 | 1.6x1.6 |
| NIRSpec (IFU) | 1.0-5.0 | 2700 | 100 | 3.0x3.0 |
| MIRI (imaging) | 5-27 | 10 | 110 | 1.4x1.9 |
| MIRI (prism) | 5-10 | 100 | 200 | 1.4x1.9 |
| MIRI (spectroscopy) | 5-27 | 3000 | 200 | 7.5x7.5 |
| NIRISS | 0.6-5.0 | 700 | 68 | 132x132 |
| JUICE MAJIS | 0.4-5.7 | ~800 | N/A | N/A |

**Table 2. JWST Observation Opportunities for Outer Planet Satellites.** Imaging opportunities during 2018-2023. In order to stay behind its sunshield, JWST can observe outer planets only near quadrature when the Sun-Earth-target angle is close to 90°. The phase angle for observations of the Jovian system and beyond is low and with minimal variation for each target.

| System | NIRCam (km/pixel) | Observable Dates |
|---|---|---|
| Jupiter | 110-120 | MAR, SEP 2019; APR, OCT 2020; MAY, NOV 2021; JUL, DEC 2022; AUG 2023 |
| Saturn | 220 | OCT, 2018; APR, OCT 2019; APR, OCT 2020; MAY, NOV 2021; MAY, NOV 2022; MAY, NOV 2023 |
| Uranus | 440-450 | JAN, JUL 2019; JAN, AUG 2020; JAN, AUG 2021; FEB, AUG 2022; FEB, AUG 2023 |
| Neptune | 670 | DEC 2018; JUN, DEC 2019; JUN, DEC 2020; JUN, DEC 2021; JUN, DEC 2022; JUN, DEC 2023 |

FIGURES

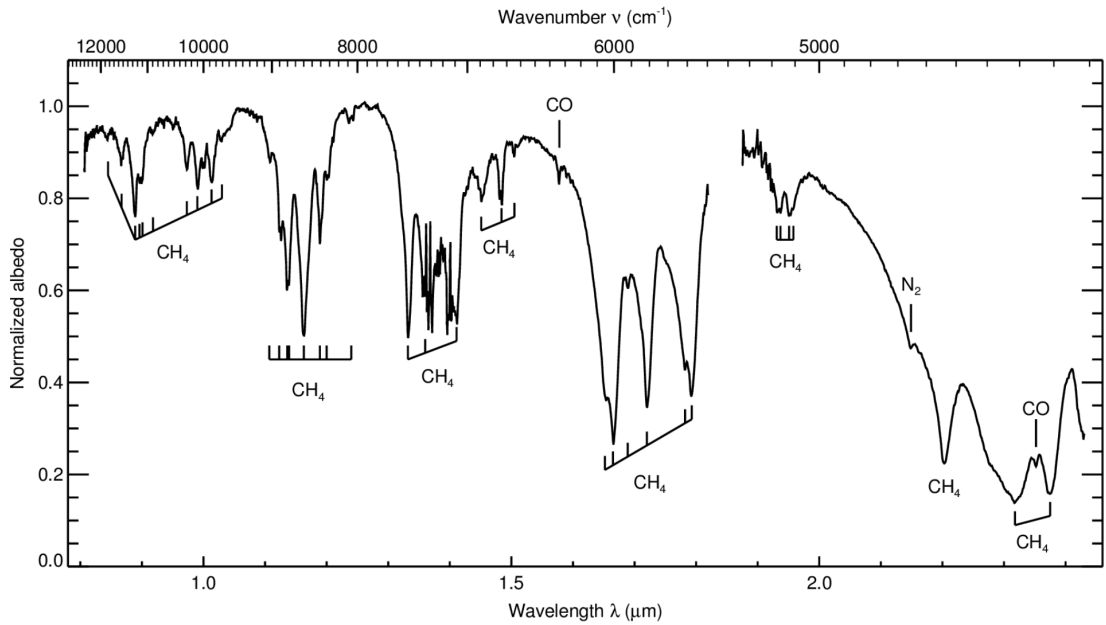

**Figure 1.** Infrared spectra of Pluto and Triton showing some of the species that can be identified in this wavelength region (after Grundy et al., 2013).

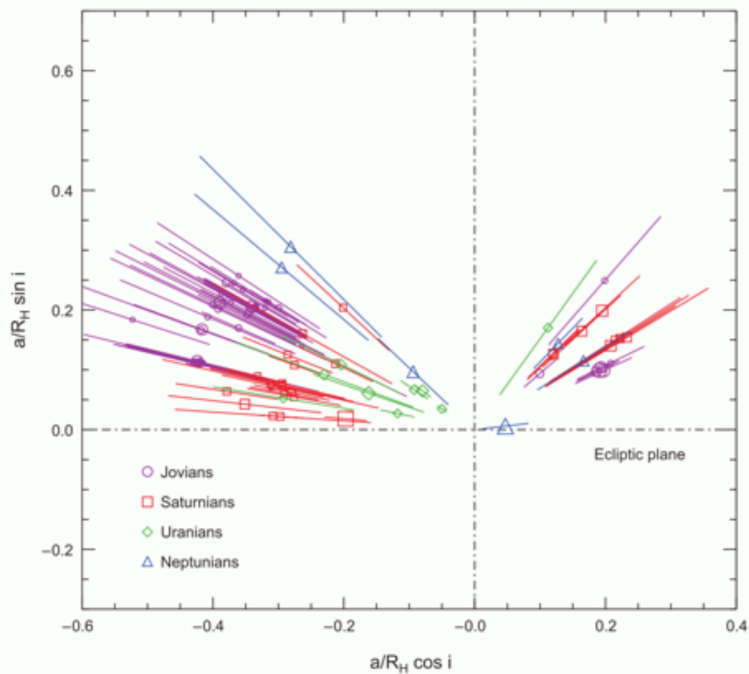

**Figure 2.** Inclinations and separations of the irregular satellites of the four giant planets showing clustering in orbital element space (from Nicholson et al., 2008).

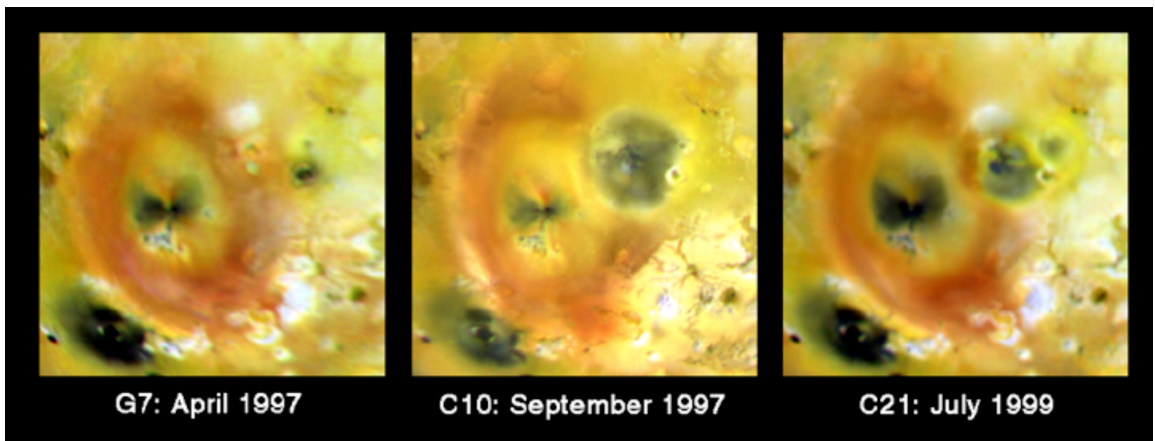

**Figure 3.** Galileo SSI images of the formation and fading of the 400 km diameter Pillan plume deposit (new black feature within the red Pele plume deposit). Changes like this could be resolved, both spatially and temporally, by JWST.

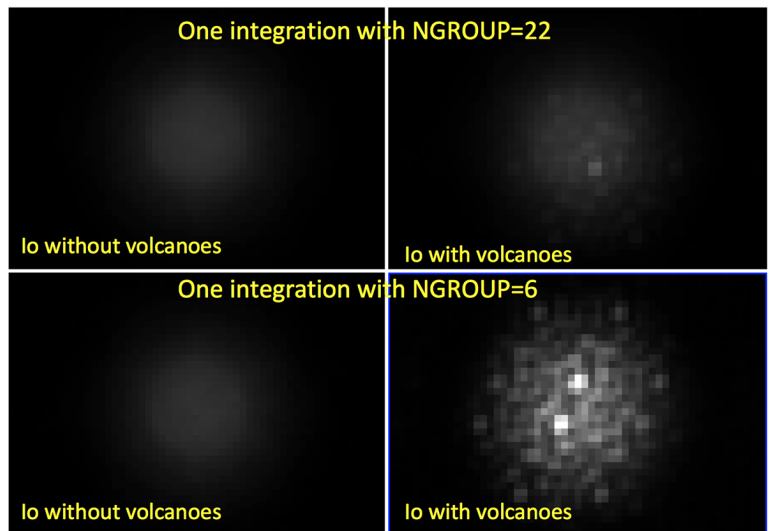

**Figure 4.** Simulated JWST observations of Io with the Non-Redundant Aperture Mask (NRM) of the NIRISS instrument. Panel a shows a model count-rate image of Io with two bright events; the pixel scale for the model image is oversampled by a factor of 10 relative to the NIRISS pixel scale. Panel b is the NRM PSF of a point-source created using WebbPSF for F430M filter, and panel c shows the convolution of the model input image with the NRM PSF (both panels us the oversampled pixels). Panel d is shows simulated data for a single exposure and at the NIRISS pixel scale (emission from Io's disk has a low signal-to-noise ratio in this image due to the few-second exposure time used). Emission from the volcanoes is readily apparent and the image illustrates the excellent spatial resolution provided by NIRISS NRM imaging.

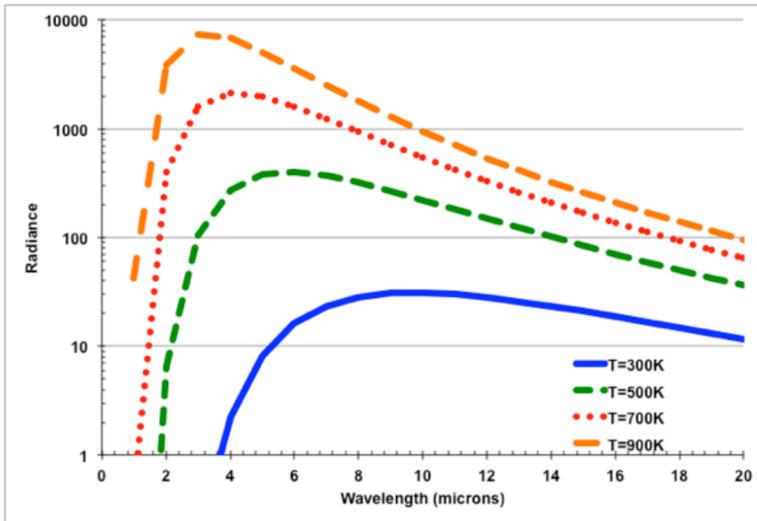

**Figure 5.** Planck Function thermal emission at temperatures relevant for active silicate volcanism on Io. Note that the response in the 1-2 micron region is extremely sensitive to the highest temperature components. Radiance is in W m$^{-2}$ μm$^{-1}$.

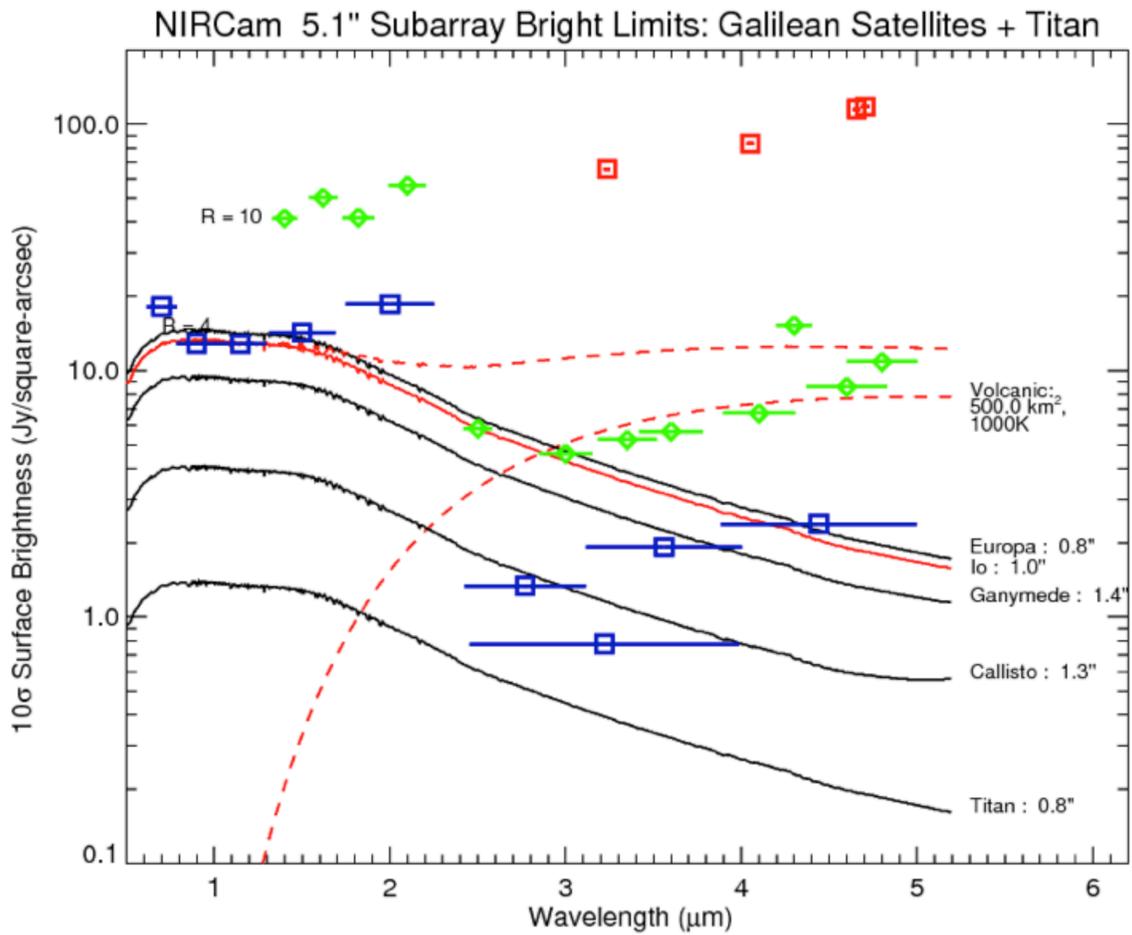

**Figure 6a.** Comparison of the predicted brightness of the largest satellites in the Solar System versus saturation limits of the JWST NIRCam. These satellites are well-resolved by JWST. The saturation limits are calculated for a small subarray (5.1" field of view) that can be read out in 0.55 seconds as opposed to other modes with longer read out and lower saturation limits. Here the reflectivity of the satellites is assumed to be a constant value (i.e., gray). Red dashed lines indicate the effect of a Loki-like eruption on the spectrum of Io. Thermal emission also affects the 4.5-5 micron spectrum of Callisto. Blue, green and red symbols show the sensitivity values in the NIRCam wide, medium and narrow-band filters (spectral resolutions of approximately 4, 10 and 100, respectively).

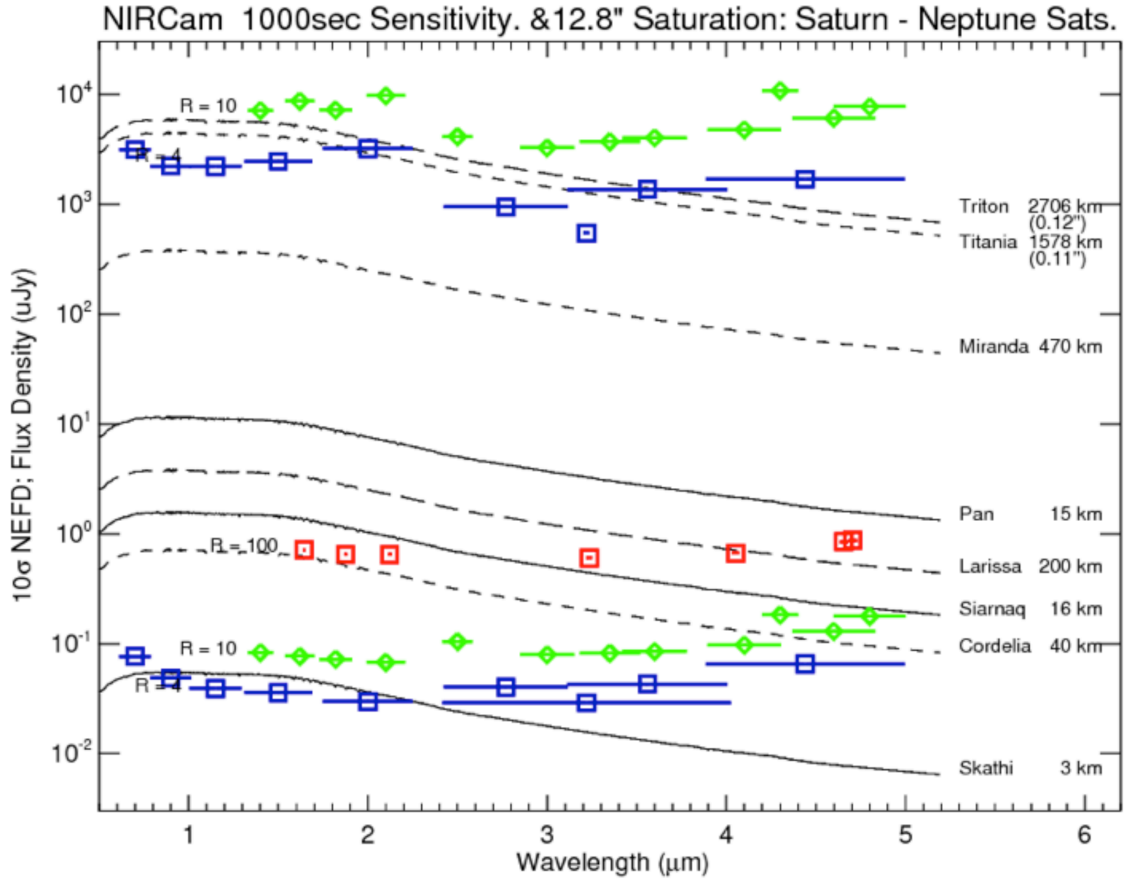

**Figure 6b.** The predicted brightness of a selection of satellites of Saturn, Uranus and Neptune is compared to saturation limits and 10-sigma point-source sensitivity for NIRCam. The saturation limits are for a 12.8" subarray that can be read out in 3.3 seconds. Sensitivity is for a 1000 second exposure. The reflectivity of the satellites is assumed to be gray. Triton and Titania will be resolved at wavelengths shortward of about 4 microns, and will have slightly lower effective flux due to being resolved. Blue, green and red symbols show the sensitivity values in the NIRCam wide, medium and narrow-band filters (spectral resolutions of approximately 4, 10 and 100, respectively).

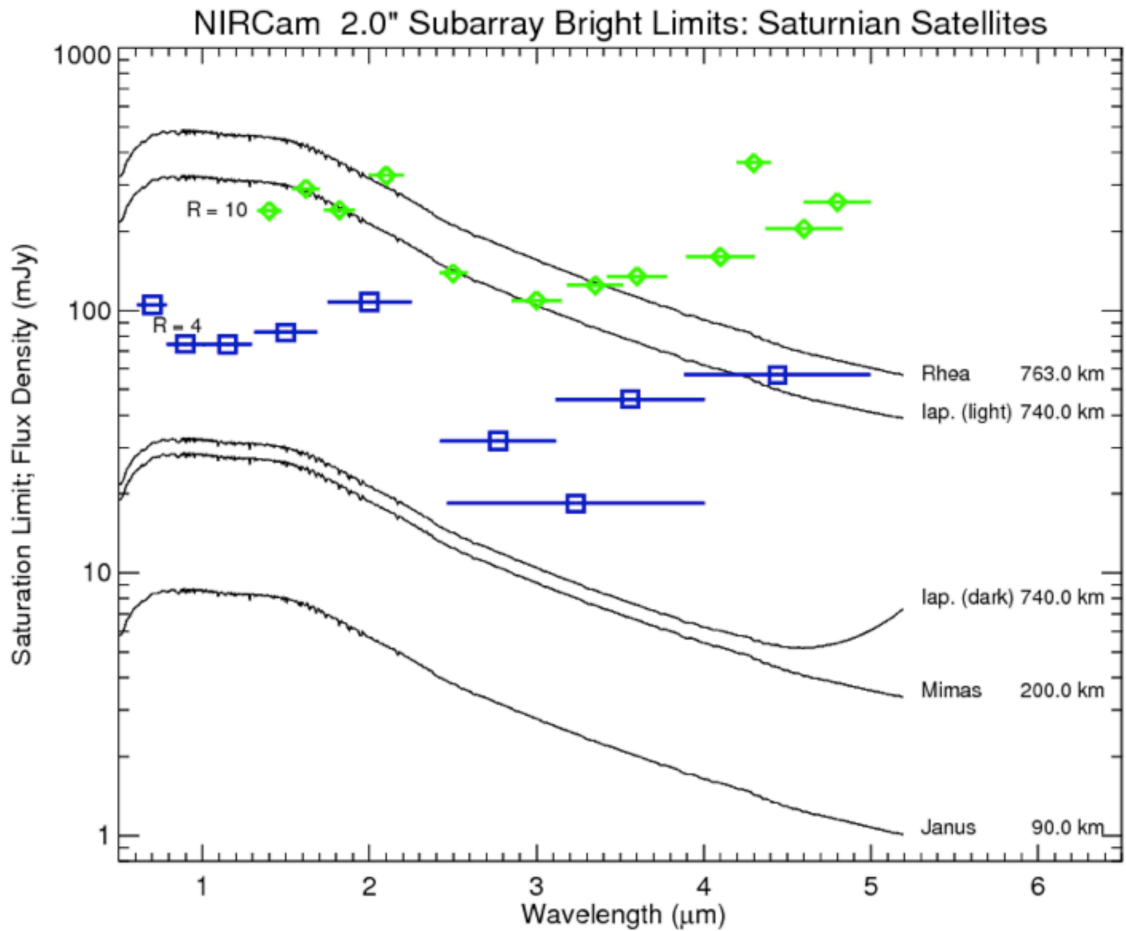

**Figure 6c.** The predicted brightness of a selection of Saturnian satellites to saturation limits for NIRCam. The saturation limits are for a 2" subarray that can be read out in 0.1 seconds. The reflectivity of the satellites is assumed to be gray. Blue, green and red symbols show the sensitivity values in the NIRCam wide, medium and narrow-band filters (spectral resolutions of approximately 4, 10 and 100, respectively).

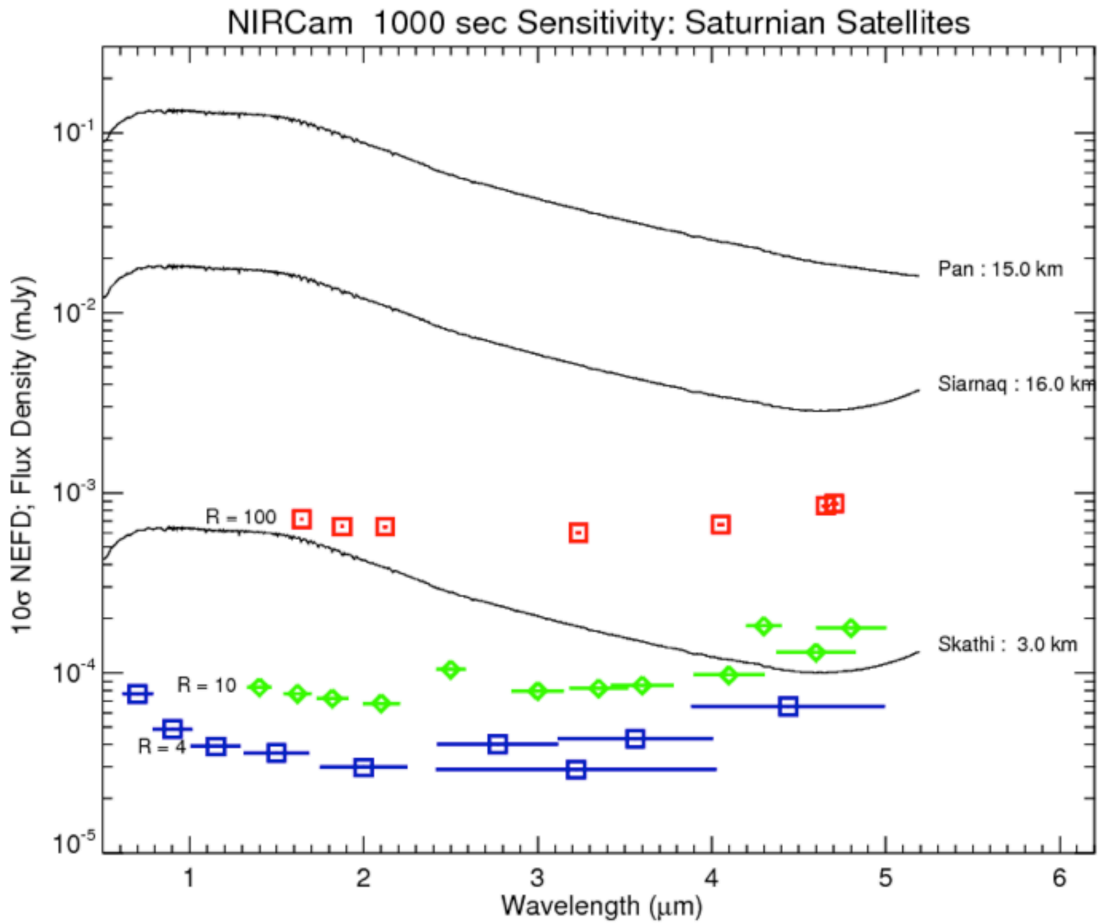

**Figure 6d.** The predicted brightness of some small satellites of Saturn 10-sigma point-source sensitivity for NIRCam. Sensitivity is for a 1000 second exposure. The reflectivity of the satellites is assumed to be gray. Blue, green and red symbols show the sensitivity values in the NIRCam wide, medium and narrow-band filters (spectral resolutions of approximately 4, 10 and 100, respectively).